\documentclass[12pt,paper,notoc]{JHEP}

\usepackage{amssymb,epsfig}

\makeatletter
\def\slash#1{{\mathpalette\c@ncel{#1}}} 
\makeatother

\newcommand{\deriv}{\stackrel{\leftrightarrow}{D}}
\newcommand{\derleft}{\stackrel{\leftarrow}{D}}
\newcommand{\derright}{\stackrel{\rightarrow}{D}}
\newcommand\wt{\widetilde}

\title{Theoretical Update of Pseudoscalar Meson Distribution
Amplitudes of Higher Twist: The Nonsinglet Case}
\author{Patricia Ball\thanks{E-mail: Patricia.Ball@cern.ch; Heisenberg
  fellow.}\\CERN--TH, CH--1211 Geneva 23, Switzerland}
\abstract{We discuss the two and three particle light-cone distribution
amplitudes (DAs) of pseudoscalar nonsinglet mesons of twist 3 and 4. Using
 nonlocal operator identities and conformal
expansion, we derive closed expressions for several DAs. 
We also include meson-mass corrections which
prove to be dominant in the twist 4 DAs of $K$ and $\eta$
mesons. Explicit parametrizations for the DAs of $\pi$, $K$ and $\eta$
mesons are given, with the numerical input parameters determined from
QCD sum rules.}
\preprint{CERN--TH/98--400\\hep-ph/9812375}

\begin{document}

\section{Introduction}
\setcounter{equation}{0}

Meson distribution amplitudes (DAs) describe
 the momentum fraction distributions
of partons in a meson, in a particular Fock state, with a fixed number
of constituents. In the standard treatment of exclusive processes in
QCD, which is due to Brodsky and Lepage \cite{BLreport},
cross sections are expanded in inverse powers of the momentum
transfer; the size of these power-suppressed corrections,
ordered by increasing twist, is determined by the convolution of a
perturbative hard scattering amplitude with a soft nonperturbative
 DA of given twist. The leading twist DA $\phi$, which describes the momentum
 distribution of the valence quarks in the meson, is related to the
 meson's Bethe--Salpeter wave function $\phi_{BS}$ by
$$
\phi(x) \sim \int^{|k_\perp| < \mu} \!\!d^2 k_\perp\,
\phi_{BS}(x,k_\perp).
$$
Here $\mu$ denotes the separation scale between perturbative and
nonperturbative regime. The study of these leading twist 2 DAs
 has attracted much attention in the literature, in
particular for the case of the $\pi$ \cite{CZreport,piWF1,piWF2}, 
but only a few
investigations are devoted to higher twist distributions, which
determine the preasymptotic behaviour of hard exclusive processes.
Higher twist DAs originate from three different sources and describe either
contributions of ``bad'' components in the wave function and in
particular of components with ``wrong'' spin projection or contributions
of transverse motion of quarks (antiquarks) in the leading twist
components or contributions of higher Fock states with additional gluons and/or
quark-antiquark pairs.

DAs of the
$\pi$ of twist 3 and 4 have been studied in \cite{BF90} in the chiral
limit, based on the techniques of nonlocal operator product
expansion and conformal expansion. In Refs.~\cite{BBKT,HB,HT98}, vector
meson DAs of twist 3 and 4 have been studied, also including
corrections in the meson-mass. In this paper we extend the analysis of
\cite{BF90} to include also terms in the meson-mass in twist 3 and 4 DAs
of pseudoscalar octet mesons. As discussed in \cite{HT98}, the
structure of these mass corrections is more complicated than for
deep-inelastic lepton-hadron scattering, where the corrections, being
 induced by kinematics, do not involve new information on dynamics and
 can be absorbed into a redefinition of the scaling variable, known
 as Nachtmann scaling \cite{N73}. The situation with exclusive decays
 is different, as matrix elements of operators containing total
 derivatives, specifically
$$\partial^2 O^{(2)}_{\mu_1\mu_2\ldots\mu_n}\quad {\rm and} \quad
\partial_{\mu_1}O^{(2)}_{\mu_1\mu_2\ldots\mu_n},$$
where $O^{(2)}$ is a leading twist operator, vanish for
forward-scattering, but do contribute to exclusive processes. 
Contributions of the first type can be taken into
account  consistently for all moments of DAs, while contributions of the
second type are more complicated and can be unravelled only order by
order in the conformal expansion.
Numerically, as  expected, these mass terms turn out to be
small for the $\pi$, but are dominant for $K$ and the octet DAs of
$\eta$. We shall not discuss the octet DAs of the $\eta'$ in this
paper. The results are of direct relevance for the
discussion of, for instance, meson transition form factors, $\gamma\gamma^*\to
\eta$, and also for $B$ meson decays into light mesons,
see e.g.\ \cite{PB98}. 

\section{Definition of Distribution Amplitudes}

Amplitudes of light-cone dominated processes involving pseodoscalar mesons
can be expressed in terms of matrix elements of gauge invariant
nonlocal operators sandwiched between the vacuum and the  meson
state, e.g.\ a matrix element over a two particle operator,
\begin{equation}
    \langle 0|\bar u(x) \Gamma [x,-x] d(-x)|\pi^-(P)\rangle,
\label{eq:1}
\end{equation}
where $\Gamma$ is a generic Dirac matrix structure
and we use the notation
$[x,y]$ for the path-ordered gauge factor along the straight line
connecting the points $x$ and $y$:
\begin{equation}
[x,y] ={\rm P}\exp\left[ig\!\!\int_0^1\!\! dt\,(x-y)_\mu
  A_\mu(tx+(1-t)y)\right].
\label{Pexp}
\end{equation}
For notational convenience, we refer explicitly to $\pi^-$ mesons. For
other (nonsinglet) mesons, one has to use
appropriate SU(3) currents, e.g.\ $1/\sqrt{2}\,\langle 0 |
\bar u \gamma_\mu\gamma_5 u - \bar d \gamma_\mu\gamma_5 d|
\pi^0\rangle$ etc.

As mentioned in the introduction, we are in particular interested in
meson-mass corrections. In contrast to vector mesons, whose mass is of
order $\Lambda_{\rm QCD}$ and nonvanishing also in the chiral limit,
  pseudoscalar meson-masses scale linearly with the sum of quark
  masses. For consistency, we will thus keep all such terms in the
  analysis of the equations of motion, but neglect terms in the
  difference of quark masses. We thus also neglect contributions in
  the DAs of $K$ mesons that are antisymmetric under the exchange of
  strange and nonstrange quark.

The asymptotic expansion of exclusive amplitudes in powers of
large momentum transfer corresponds to the expansion
of amplitudes like (\ref{eq:1}) in powers of the deviation
from the light-cone $x^2= 0$. As always in quantum field theory,
such an
expansion generates divergences and has to be understood as an
operator product expansion in terms of renormalized nonlocal
operators on the  light-cone, whose matrix elements define meson
DAs of increasing twist. To  leading logarithmic accuracy,
the coefficient functions are just taken at tree-level and the
distributions have to be evaluated at the scale $\mu^2 \sim x^{-2}$.
In this section we present the necessary expansions
and introduce a complete set of meson DAs
to twist 4 accuracy. This set is, in fact, overcomplete, and different
distributions are related to one another via the QCD equations of motion, as
 detailed in Secs.\ 5 and 6.

To facilitate the discussion of matrix elements on the light-cone, it
is convenient to introduce light-like vectors $p$ and $z$ such that
\begin{equation}
p_\mu = P_\mu-\frac{1}{2}z_\mu \frac{m^2}{pz},
\label{smallp}
\end{equation}
where $P_\mu$ is the meson momentum, $P^2=m^2$. 
We also need the projector onto the directions orthogonal to $p$ and $z$:
\begin{equation}
       g^\perp_{\mu\nu} = g_{\mu\nu} -\frac{1}{pz}(p_\mu z_\nu+ p_\nu z_\mu),
\end{equation}
and will use the notations
\begin{equation}
    a.\equiv a_\mu z^\mu, \qquad a_\ast \equiv a_\mu p^\mu/(pz),
\end{equation}
for an arbitrary Lorentz vector $a_\mu$.

We use the standard Bjorken--Drell
convention \cite{BD65} for the metric and the Dirac matrices; in particular
$\gamma_{5} = i \gamma^{0} \gamma^{1} \gamma^{2} \gamma^{3}$,
and the Levi--Civita tensor $\epsilon_{\mu \nu \lambda \sigma}$
is defined as the totally antisymmetric tensor with $\epsilon_{0123} = 1$.
The covariant derivative is defined as
$D_{\mu} = \partial_{\mu} - igA_{\mu}$.
 The dual gluon field strength
tensor is defined as $\widetilde{G}_{\mu\nu} =
\frac{1}{2}\epsilon_{\mu\nu \rho\sigma} G^{\rho\sigma}$.

We start with the two particle DAs of the $\pi$ meson. For the axial
vector operator, the light-cone expansion to twist 4 accuracy reads:
\begin{eqnarray}
\lefteqn{\langle 0 | \bar u(x)\gamma_\mu\gamma_5 d(-x)|\pi^-(P)\rangle\
  =}\hspace*{0.5cm}\nonumber\\
& = & i f_\pi P_\mu \int_0^1 du \, e^{i\xi Px} \left[ \phi_\pi(u) +
  \frac{1}{4}\, m_\pi^2 x^2 {\mathbb A}(u)\right] + \frac{i}{2}\,
  f_\pi m_\pi^2\,
  \frac{1}{Px}\, x_\mu \int_0^1 du \, e^{i\xi Px}\, 
{\mathbb B}(u).\hspace*{0.7cm}
\label{eq:T2}
\end{eqnarray}
$\phi_\pi$ is the leading twist 2 DA, $\mathbb A$ and $\mathbb B$
contain contributions from operators of twist 2, 3 and 4. 
For brevity, here and below we do not show gauge factors between the
quark and the antiquark fields; we also use the short-hand notation
$$\xi = 2u-1.$$ The decay constant $f_\pi$ is defined, as usual, as
\begin{equation}
\langle 0|\bar u(0) \gamma_{\mu}\gamma_5
d(0)|\pi^-(P)\rangle = if_{\pi}P_\mu.
\end{equation}
Numerically, one has $f_\pi = 131\,$MeV and $f_K = 160\,$MeV
\cite{PD}. For $\eta$, the situation is more complicated
due to the mixing with $\eta'$, and differing results for the coupling
to the octet current, $f_\eta^8$, are available in the
literature. We 
quote, for instance, $f_\eta^8 \approx 130\,$MeV \cite{unverhofft}
and, from a more recent analysis,
$f_\eta^8 = 159\,$MeV \cite{stech}. In the
numerical analysis we will use the ilustrative value
$f_\eta^8=130\,$MeV. As we shall see, the DAs themselves do not depend 
critically on the decay constants,
although the matrix element (\ref{eq:T2}) does.

The Lorentz invariant amplitude $\mathbb B$ can be interpreted in
terms of meson DAs, defined in terms of nonlocal operators at strictly
light-like separations, which can most conveniently be written using 
light-cone variables. For the axial vector current, 
the two particle DAs of the $\pi$ meson are defined as
\begin{eqnarray}
\lefteqn{\langle 0 | \bar u(z) \gamma_\mu\gamma_5 d(-z) |
  \pi^-(P)\rangle\ =}\hspace*{1cm}\nonumber\\
& = & i f_\pi p_\mu \int_0^1 du\, e^{i\xi pz} \, \phi_\pi(u) +
  \frac{i}{2}\, f_\pi m^2\, \frac{1}{pz}\, z_\mu \int_0^1 du \,
  e^{i\xi pz} g_\pi(u).\label{eq:2.8}
\end{eqnarray}
Comparing the above with (\ref{eq:T2}), one finds
\begin{equation}
{\mathbb B}(u) =  g_\pi(u) - \phi_\pi(u).
\end{equation}
The relation of these DAs to those defined by Braun and Filyanov,
 Ref.~\cite{BF90}, is given by 
\begin{equation}
\frac{d\phantom{u}}{du}\, g_2^{BF}(u) = -\frac{1}{2}\,
  \lim_{m_\pi^2\to 0} m_\pi^2 {\mathbb
  B}(u),\qquad g_1^{BF}(u)- \int_0^u dv g_2^{BF}(v) = 
\frac{1}{16}\, \lim_{m_\pi^2\to 0} m_\pi^2 {\mathbb A}(u).
\end{equation}
In the local limit $x_\mu\to 0$, (\ref{eq:T2}) yields the 
normalization conditions:
\begin{eqnarray*}
\int_0^1 du\, \phi_\pi(u) & = & 1,\\
\int_0^1 du\, {\mathbb B}(u) & = & 0\:\: \Longrightarrow 
\int_0^1 du\, g_\pi(u) \ = \ 1.
\end{eqnarray*}
Two more matrix elements define DAs of twist 3 \cite{BF90}:
\begin{eqnarray}
\langle 0 | \bar u(x) i\gamma_5 d(-x) | \pi^-(P)\rangle & = &
\frac{f_\pi m_\pi^2}{m_u+m_d}\, \int_0^1 du \, e^{i\xi Px}\,
\phi_p(u),\label{eq:2.11}\\
\langle 0 | \bar u(x) \sigma_{\alpha\beta}\gamma_5 d(-x) |
\pi^-(P)\rangle & = &\hspace*{9cm}\nonumber\\
\lefteqn{=\ -\frac{i}{3}\, \frac{f_\pi
  m_\pi^2}{m_u+m_d}\left\{1-\left(\frac{m_u+m_d}{m_\pi}\right)^2\right\}
  (P_\alpha x_\beta-
P_\beta x_\alpha) \int_0^1 du \, e^{i\xi Px}\,\phi_\sigma(u).}
\hspace*{4cm}\label{eq:2.12}
\end{eqnarray}
Also these two DAs are normalized to unity:
$$
\int_0^1 du\, \phi_{(p,\sigma)}(u) = 1.
$$
The normalization factor in (\ref{eq:2.12}) differs from the one
obtained in \cite{BF90} by a term of $O(m_u+m_d)\sim O(m_\pi^2)$,
which is tiny for the $\pi$, but amounts to 10\% for the $K$. This may
be of particular relevance for calculations of the $B\to K$ decay form
factor in the framework of QCD sum rules on the light-cone, e.g.\
\cite{PB98}, to which the twist 3 DAs give a sizeable contribution.

At this point we would like to comment on the numerical values to be
used for the normalization factors in (\ref{eq:2.11}) and
(\ref{eq:2.12}). Evidently, it is difficult to give precise numbers
 as long as the quark masses are not more accurately
known. To circumvent this problem, we invoke chiral 
perturbation theory (see e.g.\ \cite{pich} for a
nice introduction), which relates meson to quark masses in the following way:
define the constant $B_0$ via the (nonstrange) quark condensate:
$$
\langle 0 | \bar q q | 0 \rangle = -\frac{f_\pi^2}{2}\, B_0
$$
at the scale $\mu \approx 1\,$GeV. Then the meson-masses are given by
\begin{eqnarray}
m_\pi^2 & = & (m_u+m_d) B_0,\nonumber\\
m_K^2 & = & (m_{u,d} + m_s) B_0,\nonumber\\
m_{\eta_8}^2 & = & \frac{2}{3}\left\{ \frac{1}{2} \left( m_u +
    m_d\right) + 2 m_s\right\} B_0,
\end{eqnarray}
where we neglect small corrections in $(m_d-m_u)^2$. With the standard
value of the quark condensate, $\langle 0 |
\bar q q | 0 \rangle(1\,{\rm GeV}) = -(0.24\pm 0.01)\,{\rm GeV}^3$, 
one finds $B_0 = (1.6\pm 0.2)\,$GeV. Thus we have
$$
\frac{m_\pi^2 f_\pi^2}{m_u+m_d} = f_\pi^2 B_0 = (0.027\pm 0.003)\,{\rm
  GeV}^3.
$$
 The situation is slightly different for the $K$ and $\eta$
(which we consider as a pure octet state in this section). Proceeding like with
the $\pi$, one finds (letting $m_{u,d} = 0$):
\begin{eqnarray}
\frac{m_K^2 f_K^2}{m_s} & = & f_K^2 B_0\hspace*{0.8cm} = (0.041\pm0.005)\,{\rm
  GeV}^3,\nonumber\\
\frac{m_\eta^2 (f^8_\eta)^2}{m_s} & = & \frac{4}{3}\, (f^8_\eta)^2 B_0
  = (0.045\pm0.006)\,{\rm GeV}^3,\label{eq:x}
\end{eqnarray}
where one might worry, however, that the constant $B_0$ be
affected by SU(3) violation. Using the actual values for the meson
masses and $m_s(1\,{\rm GeV}) = 150\,$MeV, one finds 
$$
\frac{m_K^2 f_K^2}{m_s}  = 0.042\,{\rm GeV}^3,\quad 
\frac{m_\eta^2 (f^8_\eta)^2}{m_s} = 0.042 \,{\rm GeV}^3,
$$
which  is in good agreement with
the results from chiral perturbation theory.

Let us now define the three particle DAs. To twist 3 accuracy, there
is only one:
\begin{eqnarray}
\lefteqn{\langle 0 | \bar u(z) \sigma_{\mu\nu}\gamma_5
  gG_{\alpha\beta}(vz) d(-z)| \pi^-(P)\rangle\ =}\hspace*{0.5cm}\nonumber\\
& = & i\,\frac{f_\pi m_\pi^2}{m_u+m_d} \left(p_\alpha p_\mu
  g_{\nu\beta}^\perp - p_\alpha p_\nu
  g_{\mu\beta}^\perp - p_\beta p_\mu g_{\nu\alpha}^\perp + p_\beta
  p_\nu g_{\alpha\mu}^\perp \right) {\cal T}(v,pz) +
  \dots,\label{eq:3pT3}
\end{eqnarray}
where the ellipses stand for Lorentz structures of twist 5 and higher
and where we used the following short-hand notation for the integral
defining the three particle DA:
\begin{equation}
{\cal T}(v,pz) = \int {\cal D}\underline{\alpha} \, e^{-ipz(\alpha_u
  -\alpha_d + v\alpha_g)} {\cal T}(\alpha_d,\alpha_u,\alpha_g).
\end{equation}
Here $\underline{\alpha}$ is the set of three momentum fractions
$\alpha_d$ ($d$ quark), $\alpha_u$ ($u$ quark) and $\alpha_g$
(gluon). The integration measure is defined as
$$
\int {\cal D}\underline{\alpha} = \int_0^1 d\alpha_d d\alpha_u
d\alpha_g \delta(1-\alpha_u-\alpha_d-\alpha_g).
$$
There are also four three particle DAs of twist 4, defined as
\begin{eqnarray}
\lefteqn{\langle 0 | \bar u(z)\gamma_\mu\gamma_5
gG_{\alpha\beta}(vz)d(-z)|\pi^-(P)\rangle\ =}\hspace*{0.5cm}\nonumber\\
& = & p_\mu (p_\alpha z_\beta - p_\beta z_\alpha)\, \frac{1}{pz}\, f_\pi
m_\pi^2 {\cal A}_\parallel(v,pz) + (p_\beta g_{\alpha\mu}^\perp -
p_\alpha g_{\beta\mu}^\perp) f_\pi m_\pi^2 {\cal A}_\perp(v,pz),\\
\lefteqn{\langle 0 | \bar u(z)\gamma_\mu i
g\widetilde{G}_{\alpha\beta}(vz)d(-z)|\pi^-(P)\rangle\
=}\hspace*{0.5cm}\nonumber\\
& = & p_\mu (p_\alpha z_\beta - p_\beta z_\alpha)\, \frac{1}{pz}\, f_\pi
m_\pi^2 {\cal V}_\parallel(v,pz) + (p_\beta g_{\alpha\mu}^\perp -
p_\alpha g_{\beta\mu}^\perp) f_\pi m_\pi^2 {\cal V}_\perp(v,pz).
\end{eqnarray}

A short synopsis of the various light-cone projections of the
three-particle matrix elements and their relation to DAs is
given in Table~\ref{tab:LC}.

\TABLE{
\begin{tabular}{|c|lc|lcc|}
\hline
Twist & $(\mu\nu\alpha\beta)$
& $\bar\psi \sigma_{\mu\nu}\gamma_5\widetilde{G}_{\alpha\beta}
\psi$ & $(\mu\alpha\beta)$ & $\bar\psi \gamma_\mu\gamma_5 G_{\alpha\beta} \psi$
& $\bar\psi \gamma_\mu\wt{G}_{\alpha\beta}\psi$\\ \hline
3 & $\cdot\perp \cdot\perp$ & ${\cal T}$ & & &\\
4 & & & $\cdot\cdot *$ & ${\cal A}_\parallel$ & ${\cal V}_\parallel$\\ 
 & & & $\perp\perp\cdot$  & ${\cal A}_\perp$ & ${\cal V}_\perp$\\ \hline
\end{tabular}
\caption{
Identification of three-particle DAs with
projections onto different light-cone components of 
nonlocal operators. For example, $\perp\perp\cdot$ refers to
$\bar\psi \gamma_\perp\gamma_5 G_{\perp\cdot} \psi$.
}\label{tab:LC}
}

For completeness, let us mention that also four particle twist 4
DAs exist, corresponding to contributions of
Fock states with two gluons or an additional $q \bar q$ pair.
Such distributions will not be considered in this paper for two
reasons: first, it is well known \cite{SV82} that four particle
twist 4 operators do not allow the    factorization of  vacuum condensates
such as $\langle \bar\psi \psi\rangle$, $\langle G^2\rangle$.
Because of this,  their matrix elements cannot be estimated reliably by
existing methods (e.g.\ QCD sum rules), although they are
generally  expected to be small.
Second, and more importantly, the four particle
distributions decouple from the QCD equations of motion in the two lowest
conformal partial waves. To this accuracy, therefore, it is consistent
to put them to zero. Vice versa, nonvanishing four particle
distributions necessitate the inclusion of higher conformal spin
corrections to distributions with less particles, which are beyond
the approximation adopted in this paper.

\section{Conformal Partial Wave Expansion and Equations of Motion}
\setcounter{equation}{0}

The aim of this paper is to express the DAs defined in the previous
section in a model-independent way by a minimal number of
nonperturbative parameters. The one key ingredient in solving this task is
the use of the QCD equations of motion which will allow us to
reveal interrelations between the different DAs of a given
twist. Nonlocal operators on or near the light-cone can conveniently
be treated in the framework of the string-operator technique
developped by Balitskii and Braun \cite{BB89}. In
the present context, we need the following nonlocal 
operator identities \cite{HT98}:
\begin{eqnarray}
\frac{\partial}{\partial x_\mu}\, \bar u(x)\gamma_\mu\gamma_5 d(-x)
& = &{} - i \int_{-1}^1 dv\, v \bar u (x) x_\alpha
gG_{\alpha\mu}(vx) \gamma_\mu\gamma_5 d(-x)\nonumber\\
& &  + (m_u-m_d) \bar u(x) i\gamma_5 d(-x),\label{eq:oprel1}\\
\partial_\mu \{\bar u(x)\gamma_\mu\gamma_5 d(-x)\}
& = & {} - i\int_{-1}^1 dv\, \bar u(x) x_\alpha
gG_{\alpha\mu}(vx) \gamma_\mu\gamma_5 d(-x)\nonumber\\
& & {} + (m_u+m_d)\bar u(x)
i\gamma_5 d(-x),\label{eq:oprel2}\\
\partial_\mu \bar u(x) \sigma_{\mu\nu}\gamma_5 d(-x) & = &
-i\,\frac{\partial\phantom{x_\nu}}{\partial x_\nu} \,\bar u(x)\gamma_5
d(-x) + \int_{-1}^1 dv\, v \bar u(x)
x_\rho gG_{\rho\nu}(vx)\gamma_5d(-x)\nonumber\\
& & {} -i\int_{-1}^1 dv\, \bar u(x) x_\rho gG_{\rho\mu}(vx)
\sigma_{\mu\nu} \gamma_5d(-x)\nonumber\\
& &  + (m_d-m_u) \bar u(x) \gamma_\nu\gamma_5 d(-x),\label{eq:oprel3}\\
\frac{\partial\phantom{x_\nu}}{\partial x_\mu} \,\bar u(x)
\sigma_{\mu\nu}\gamma_5 d(-x) & = & -i\partial_\nu \bar u(x)
\gamma_5d(-x) +
\int_{-1}^1 dv\, \bar u(x) x_\rho gG_{\rho\nu}(vx)\gamma_5d(-x)\nonumber\\
& & {} -
i\int_{-1}^1 dv\, v \bar u(x) x_\rho gG_{\rho\mu}(vx)
\sigma_{\mu\nu}\gamma_5 d(-x)\nonumber\\
& & {}- (m_u+m_d) \bar u(x) \gamma_\nu \gamma_5 d(-x).
\label{eq:oprel4}
\end{eqnarray}
Here $\partial_\mu$ is the total derivative defined as 
$$
\partial_\mu \left\{ \bar u(x)\Gamma d(-x)\right\} \equiv
\left.\frac{\partial}{\partial y_\mu}\,\left\{ \bar u(x+y) [x+y,-x+y]
    \Gamma d(-x+y)\right\}\right|_{y\to 0}.
$$
By taking matrix elements of the above relations between the vacuum
und the $\pi^-$ meson state, one obtains exact integral
representations for those DAs that are not dynamically independent.

The other key ingredient in our approach is the use of conformal
expansion \cite{B+,BF90} which, analogously to
partial wave decomposition in quantum mechanics, allows one to separate
transverse and longitudinal variables in the wave function.  The
dependence on transverse coordinates is represented as scale-dependence
of the relevant operators and is governed by
renormalization-group equations, the dependence on the longitudinal
momentum fraction is described in terms of irreducible
representations of the corresponding symmetry group, the collinear
conformal group SL(2,$\mathbb R$). The conformal partial wave expansion is
explicitly consistent with the equations of motion since the latter are
not renormalized. The expansion thus makes maximum use of the symmetry
of the theory in order to simplify the dynamics. 

To construct the conformal expansion for an arbitrary multi-particle
distribution, one first has to decompose each constituent field into
components with fixed Lorentz spin projection onto the
light-cone. Each such component has conformal spin
$$
j=\frac{1}{2}\, (l+s),
$$
where $l$ is the canonical dimension  and $s$ the (Lorentz) spin
projection. In particular, $l=3/2$ for quarks and $l=2$ for gluons.
The  quark field is decomposed as $\psi_+ \equiv
(1/2)\slash{z}\slash{p}\psi$ and $\psi_-=
(1/2)\slash{p}\slash{z}\psi$ with
spin projections $s=+1/2$ and $s=-1/2$, respectively. For the gluon
field strength there are three possibilities:
 $G_{\cdot\perp}$ corresponds to $s=+1$,
$G_{*\perp}$ to $s=-1$ and both
$G_{\perp\perp}$ and $G_{\cdot *}$ correspond to $s=0$.

Multi-particle states built of fields with definite Lorentz spin
projection can be expanded in
irreducible unitary representations of SL(2,$\mathbb R$) 
with increasing conformal spin.
The explicit expression for the DA
of a $m$-particle state with the lowest possible conformal spin
 $j=j_1+\ldots+j_m$, the so-called asymptotic DA, is
\begin{equation}
\phi_{as}(\alpha_1,\alpha_2,\cdots,\alpha_m) =
\frac{\Gamma(2j_1+\cdots +2j_m)}{\Gamma(2j_1)\cdots \Gamma(2j_m)}\,
\alpha_1^{2j_1-1}\alpha_2^{2j_2-1}\ldots \alpha_m^{2j_m-1}.
\label{eq:asymptotic}
\end{equation}
Here $\alpha_k$ are the corresponding momentum fractions.
This state is nondegenerate and cannot mix with
other states because of conformal symmetry.
Multi-particle irreducible representations with higher spin
$j+n,n=1,2,\ldots$,
are given by  polynomials of $m$ variables (with the constraint
$\sum_{k=1}^m \alpha_k=1$ ), which are orthogonal over
 the weight-function (\ref{eq:asymptotic}).

In particular, for the leading twist 2 DA $\phi_\pi$ defined in
(\ref{eq:T2}), the expansion goes in Gegenbauer polynomials:
\begin{equation}
\phi_\pi(u,\mu^2) = 6 u (1-u) \left( 1 + \sum\limits_{n=1}^\infty
  a_{2n}(\mu^2) C_{2n}^{3/2}(2u-1)\right).
\end{equation}
To leading logarithmic accuracy, the (nonperturbative) Gegenbauer
moments $a_n$ renormalize multiplicatively with 
$$
a_n(Q^2) = L^{\gamma_n/b}\, a_n(\mu^2),
$$
where $L\equiv \alpha_s(Q^2)/\alpha_s(\mu^2)$, $b=(11N_c-2N_f)/3$, and
the anomalous dimension $\gamma_n$ is given by
$$
\gamma_n =  4C_F \left(\psi(n+2) + \gamma_E - \frac{3}{4} -
  \frac{1}{2(n+1)(n+2)} \right).
$$
In this paper, we work to next-to-leading order in conformal spin and
thus truncate\footnote{Note that a thorough discussion of
  the shape of the $\pi$ DA of leading twist necessitates the
  inclusion of higher terms in the conformal expansion. In this paper,
  however, we concentrate on higher twist DAs which constitute
  {\em corrections} to the leading twist DAs that are suppressed by powers
  of the characteristic momentum transfer in hard reactions, so we
  feel justified in neglecting higher order conformal corrections to these
  corrections.} the above expansion of $\phi_\pi$ after the term in
$n=1$. For the $\pi$, the corresponding Gegenbauer moment was
determined in e.g.\ \cite{piWF1} from QCD sum rules, for $K$, we use the
value determined in \cite{PB98}:
\begin{equation}
a_2^\pi(1\,{\rm GeV}) = 0.44, \quad a_2^K(1\,{\rm GeV}) = 0.2, \quad
a_2^\eta(1\, {\rm GeV}) = 0.2.
\end{equation}
The value for $\eta$ is new and follows from an analysis of the QCD sum rule in
\cite{piWF1} by fixing the continuum threshold $s_0$ to reproduce
$f_\eta^8 = 130\,$MeV.

\section{Meson-Mass Corrections}
\setcounter{equation}{0}

The structure of meson-mass corrections in inclusive processes is in
general more complicated than that of target-mass corrections in deep
inelastic scattering, which can be resummed using the
Nachtmann variable \cite{N73}. The terms entering the
Nachtmann variable are just the subtracted traces of the leading twist
forward-scattering matrix element, 
which can also for exclusive processes be summed to all
orders. Let us illustrate this point with the aid of the two-point
correlation function of scalar fields:
$$
\langle 0 | \phi(x)\phi(-x) | P\rangle = \int_0^1 du \, e^{i\xi Px} 
\left[ \psi(u) + \frac{1}{4}\,
  x^2 m^2 \psi_2(u) + O(x^4)\right]
$$
Here $\psi$ is the leading twist DA; $\psi_2$ receives contributions
from both the subtraction of traces in the leading twist operator,
also dubbed ``kinematical'' corrections, and
from intrinsic ``dynamical'' higher twist corrections.
The subtraction of traces can be done on the operator level by making
use of the condition \cite{BB89}
\begin{equation}
\frac{\partial^2}{\partial x_\alpha
  \partial x_\alpha}\, \left[ \phi(x)\phi(-x)\right]_{\rm l.t.} =0,
\end{equation}
which translates into the condition that all local operators arising
in the Taylor expansion be traceless. A formal solution is \cite{BB89}
\begin{eqnarray*}
\lefteqn{\left[\phi(x)\phi(-x)\right]_{\rm l.t.}\ =}\hspace*{0.5cm}\\
& = &  \phi(x)\phi(-x) +
\sum_{n=1}^\infty \int_0^1 dt \,t\left(-\frac{1}{4}\, x^2\right)^n
\left[ t (1-t)\right]^{n-1} \left[\frac{\partial^2}{t^2 \partial x_\alpha
    \partial x_\alpha}\right]^n \phi(tx)\phi(-tx).
\end{eqnarray*}
To order $x^2$, one thus has 
\begin{eqnarray}
\lefteqn{\phi(x)\phi(-x)\ =}\hspace*{0.5cm}\nonumber\\
& = &  \left[\phi(x)\phi(-x)\right]_{\rm l.t.} -
\frac{1}{4}\, x^2 \int_0^1 dt \, t\, \partial^2
 [\phi(tx)\phi(-tx)]_{\rm l.t.} + \:{\rm interaction\ terms}\: + 
O(x^4).\nonumber\\[-10pt]\label{eq:4.2}
\end{eqnarray}
Note that $\partial^2[\ldots]_{\rm l.t.}$ is a higher twist operator.
Taking a forward-scattering matrix element, $\langle P | \dots | P\rangle$,
the second term on the right-hand side vanishes, and all
mass corrections arise from subtracting traces in the leading twist
matrix element. This is the source of the Nachtmann corrections.
In the following we will calculate the corresponding  
 matrix element for exclusive processes,
\begin{equation}\label{eq:lt}
\langle 0 | [\phi(x)\phi(-x)]_{\rm l.t.} | P\rangle,
\end{equation}
{\em exactly}, i.e.\ summing up all terms in $x^2$. 

It is clear that (\ref{eq:lt}) can only depend on the
leading twist DA $\psi(u)$. Writing (\ref{eq:lt}) as a Taylor-series
over the moments of $\psi$, we can subtract the traces explicitly for each term
in the expansion, which yields ($P^2 = m^2$):
\begin{equation}
\langle 0 | [\phi(x)\phi(-x)]_{\rm l.t.} | P\rangle = \sum\limits_{n=0}^\infty
\frac{i^n}{n!}\, \langle\langle O_n\rangle\rangle \left( \frac{1}{4}\,
  m^2 x^2\right)^{n/2} U_n( Px/\sqrt{m^2 x^2}).
\label{eq:2}
\end{equation}
Here $U_n$ are the Chebyshev polynomials
$$
U_n(x) = \sum\limits_{j=0}^{\left[\frac{n}{2}\right]} (-1)^j \left( 
\begin{array}{c} n-j \\ j\end{array}
\right) (2x)^{n-2j},
$$
and $\langle\langle O_n\rangle\rangle$ are the moments of $\psi$:
$$
\langle\langle O_n\rangle\rangle = \int_0^1 du\, \xi^n \,\psi(u).
$$
Using the generating function of $U_n$, it proves possible to sum
(\ref{eq:2})
explicitly (see also \cite{BB91}):
\begin{eqnarray}
\lefteqn{\langle 0 | [\phi(x)\phi(-x)]_{\rm l.t.} | P\rangle\
 =}\hspace*{0.5cm}\nonumber\\
& = & \int_0^1 du
\,\psi(u) \frac{1}{\sqrt{(Px)^2 - m^2 x^2}} \frac{d\phantom{u}}{du}
\left[ e^{i\xi Px/2} \sin (\xi \sqrt{(Px)^2 - m^2 x^2}/2) \right].
\end{eqnarray}
In the spirit of Nachtmann, we would like to absorb all terms in $m^2$
into a new scaling variable $Px\to (Px)'$. Although we did not succeed
in finding such a variable, the above expression can be simplified 
considerably by introducing
$$
2 (Px)' \equiv Px + \sqrt{(Px)^2 - m^2 x^2},
$$
so that
\begin{eqnarray}
\lefteqn{\langle 0 | [\phi(x)\phi(-x)]_{\rm l.t.} | P\rangle\
  =}\hspace*{0.5cm}\nonumber \\
& = & \int_0^1 du\, \frac{\psi(u)}{1-\frac{m^2 x^2}{4 (Px)'^2}} \left
  ( \exp\{i\xi
  (Px)'\} - \frac{m^2 x^2}{4 (Px)'^2} \exp\left\{ i\xi (Px)' \,
  \frac{m^2 x^2}{4 (Px)'^2} \right\}\right).\label{eq:4.4}
\end{eqnarray}
Expanding to $O(x^2)$, this can be written as
$$
\langle 0 | [\phi(x)\phi(-x)]_{\rm l.t.} | P\rangle\
 =  \int_0^1 dt \int_0^1 du\, \left[ e^{i\xi Px} + \frac{1}{4}\, m^2
  x^2 t\, \xi^2 e^{i\xi Px t}\right] \psi(u),
$$
and combining with (\ref{eq:4.2}), we get
$$
\langle 0 | \phi(x)\phi(-x)] | P\rangle\
 =  \int_0^1 dt \int_0^1 du\, \left[ e^{i\xi Px} + \frac{1}{4}\, m^2
  x^2 t (1+\xi^2) e^{i\xi Px t}\right] \psi(u).
$$
This means that both sources of mass corrections, the subtraction of
traces in the leading twist matrix element and the higher twist
operator containing total derivatives, act in the same direction and
thus enlarge the mass correction terms.
We will observe the same effect, enlarged mass corrections, also in QCD.

\section{Twist 3 Distribution Amplitudes}
\setcounter{equation}{0}

The twist 3 DAs of the $\pi$ have already been studied in
\cite{BF90}. Here we extend this study by including terms in
$\rho_\pi^2 \equiv (m_u+m_d)^2/m_\pi^2 \sim O(m_\pi^2)$. 

To next-to-leading order in conformal spin, the only three particle DA
$\cal T$ gets expanded as
\begin{equation}
{\cal T}(\underline{\alpha}) = 360\eta_3 \alpha_u\alpha_d\alpha_g^2
\left\{ 1 + \omega_3\, \frac{1}{2}\left( 7\alpha_g-3\right)\right\}.
\end{equation}
$\eta_3$ is defined as
\begin{eqnarray}
\lefteqn{\langle 0 | \bar u \sigma_{\mu\nu}\gamma_5 gG_{\alpha\beta}
d|\pi^-\rangle\ =}\hspace*{0.5cm}\nonumber\\
& = & if_\pi \eta_3 \,\frac{m_\pi^2}{m_u+m_d} \left(
 P_\alpha P_\mu g_{\nu\beta} - P_\alpha P_\nu
  g_{\mu\beta} - P_\beta P_\mu g_{\nu\alpha} + P_\beta
  P_\nu g_{\alpha\mu} \right),
\end{eqnarray}
and $\omega_3$ is defined as
\begin{eqnarray}
\lefteqn{\langle 0 | \bar u \sigma_{\mu\xi} \gamma_5 [iD_\mu,gG_{\alpha\xi}] d
- \frac{3}{7}\, i\partial_\beta \bar u \sigma_{\mu\xi} \gamma_5
gG_{\alpha\xi} d | \pi^-\rangle \ =}\hspace*{0.5cm}\nonumber\\
& = & i \,\frac{f_\pi m^2}{m_u+m_d} \, 2 P_\alpha P_\beta P_\mu
\,\frac{3}{28}\, \eta_3 \omega_3 + O({\rm higher\ twist}).
\end{eqnarray}
In the notations of \cite{BF90}:
$$\eta_3 \equiv R = \frac{f_{3\pi}}{f_\pi}\, \frac{m_u+m_d}{m_\pi^2},
\qquad \omega_3 \equiv \omega_{10}.$$
These parameters are scale-dependent with ($C_A = N_c$)
\begin{eqnarray*}
\eta_3(Q^2) & = & L^{\gamma_3^\eta/b} \eta_3(\mu^2),\quad
\gamma_3^\eta = \frac{16}{3}\, C_F + C_A,\\
\omega_3(Q^2) & = & L^{\gamma_3^\omega/b} \omega_3(\mu^2),\quad
\gamma_3^\omega = -\frac{25}{6}\, C_F + \frac{7}{3}\,C_A.
\end{eqnarray*}
Numerical values are obtained from QCD sum rules \cite{BF90,CZZ} and
collected in Table~\ref{tab:2}.

\FIGURE[b]{\epsfig{file=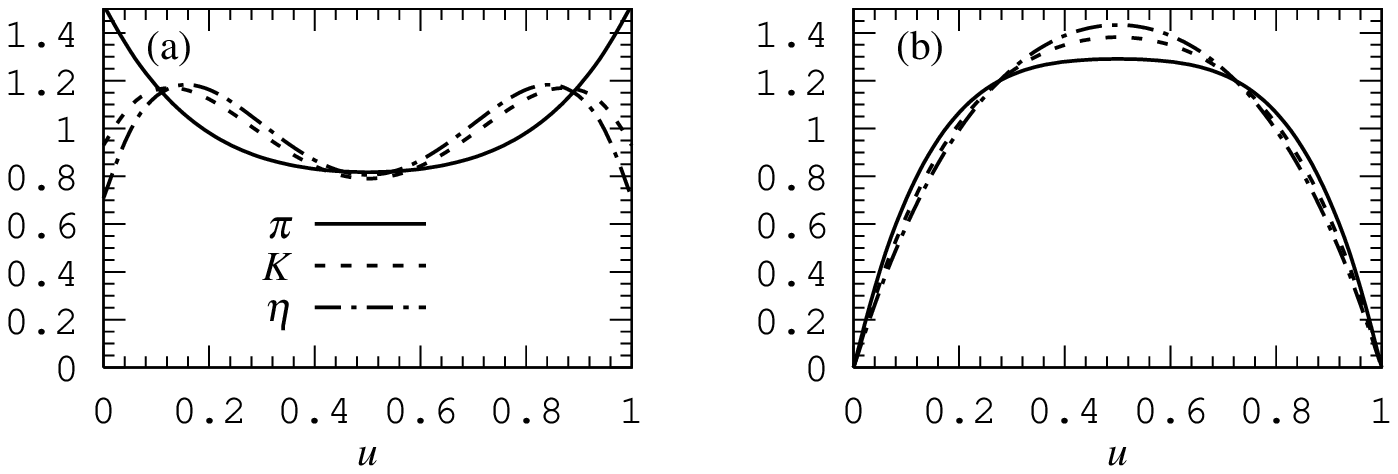}\caption{The two particle DAs of twist
    3: $\phi_p(u)$ (a) and $\phi_\sigma(u)$ (b).}\label{fig:1}}

The two particle DAs $\phi_{p,\sigma}$ are determined by ${\cal T}$
 and $\rho_\pi^2 \phi_\pi$. As an analysis of the matrix elements of
 the exact operator relations in Sec.~3 leads to integro-differential
 equations that cannot be solved in a closed form, we prefer to
 perform an analysis of moments:
\begin{eqnarray}
M_n^p & = & \delta_{n0} + \frac{n-1}{n+1}\, M^p_{n-2} + 2 (n-1)
M^{{\cal T}_1}_{n-2} + \frac{2(n-1)(n-2)}{n+1}\, M^{{\cal T}_2}_{n-3}
- \rho_\pi^2 \,\frac{n-1}{n+1}\, M_{n-2}^\phi,\nonumber\\
M_n^\sigma & = & \delta_{n0} + \frac{n-1}{n+3}\, M^\sigma_{n-2} + \frac{6
  (n-1)}{n+3} \, M^{{\cal T}_1}_{n-2} + \frac{6n}{n+3}\, M^{{\cal T}_2}_{n-1}
- \rho_\pi^2 \,\frac{3}{n+3}\, M_n^\phi.
\end{eqnarray}
Here we use the notation $M_n^P = \int_0^1 du\, \xi^n \phi_P(u)$ and
the functions
$$
\phi_{{\cal T}_1} = \int_0^u d\alpha_d \int_0^{\bar u} d\alpha_u
\,\frac{2}{\alpha_g} \,{\cal T}(\underline{\alpha}), \quad 
\phi_{{\cal T}_2} = \int_0^u d\alpha_d \int_0^{\bar u} d\alpha_u
\,\frac{2}{\alpha_g^2} \,(\alpha_d-\alpha_u-\xi)\,{\cal T}(\underline{\alpha}).
$$
Except for the new terms in $\rho_\pi^2$, the relations for moments
agree with those obtained in \cite{BF90}.

Conformal expansion imposes that $\phi_p$ gets expanded in Gegenbauer
polynomials $C_n^{1/2}$ and $\phi_\sigma$ in $C_n^{3/2}$
\cite{BF90}. {}From the recursion relations for moments we find:
\begin{eqnarray}
\phi_p(u) & = & 1 + \left(30\eta_3 -\frac{5}{2}\, \rho_\pi^2\right) 
C_2^{1/2}(\xi) + \left(- 3 \eta_3 \omega_3-\frac{27}{20}\, \rho_\pi^2
  - \frac{81}{10}\, \rho_\pi^2 a_2\right)  C_4^{1/2}(\xi),\nonumber\\
\phi_\sigma(u) & = & 6u(1-u) \left\{ 1 +
    \left(5\eta_3 -\frac{1}{2}\,\eta_3\omega_3 - \frac{7}{20}\,
      \rho_\pi^2 - \frac{3}{5}\,\rho_\pi^2 a_2 \right)
    C_2^{3/2}(\xi)\right\}.
\label{eq:3part}
\end{eqnarray}
\TABLE{
\begin{tabular}{|c|ccc|}
\hline
& $\pi$ & $K$ & $\eta$\\ \hline
$a_2$ & 0.44 & 0.2 & 0.2\\
$\eta_3$ & 0.015 & 0.015 & 0.013\\
$\omega_3$ &  $-3$ & $-3$ & $-3$\\
$\mu^2\,[{\rm GeV}]^2$ & 0.0077 & 0.096 & 0.12\\ \hline
\end{tabular}
\caption{Input parameters for twist 3  DAs,
calculated from QCD sum rules.
The accuracy is about 30\%. Renormalization scale is
1~GeV.}
\label{tab:2}}
In Fig.~\ref{fig:1} we plot the two two particle DAs of twist 3 for
$\pi$, $K$ and $\eta$ mesons. Evidently, the effect of mass
corrections is not negligible and for $\phi_p$ even modifies the shape
of the DA near the endpoints, which is due to the dependence of the
coefficient of $C_4^{1/2}$ on $\rho_\pi^2$. We would like to recall,
however, that the above parametrizations are to be understood in the
sense of (mathematical) distributions rather than as models that are
valid point by point, and that they are always intended to be
convoluted with smooth perturbative scattering amplitudes, which
in particular will smooth out the effect of neglected 
higher order terms in the conformal expansion. 

\section{Twist 4 Distribution Amplitudes}
\setcounter{equation}{0}

In this section we repeat the analysis of twist 4 DAs performed in
\cite{BF90} in a more systematic way and extend it by including 
 mass correction terms.

 Due to G-parity, in the chiral limit, the DAs ${\cal
  A}_\parallel$ and ${\cal A}_\perp$ are antisymmetric under the
exchange of $\alpha_d$ and $\alpha_u$, whereas ${\cal V}_\parallel$
and ${\cal V}_\perp$ are symmetric; contributions of ``wrong''
  G-parity give rise to asymmetric contributions to the two particle
  DAs of $K$ and are neglected in the following. The distributions ${\cal
  A}_\parallel$ and ${\cal V}_\parallel$ correspond to the light-cone
projection $\gamma_\cdot G_{\cdot *}$ (see Table~\ref{tab:LC}) and
have the conformal expansion
\begin{eqnarray}
{\cal V}_\parallel(\underline{\alpha}) & = & 120
\alpha_u\alpha_d\alpha_g ( v_{00} + v_{10} (3\alpha_g-1) +
\dots),\nonumber\\
{\cal A}_\parallel(\underline{\alpha}) & = & 120
\alpha_u\alpha_d\alpha_g ( 0 + a_{10} (\alpha_d-\alpha_u) +
\dots),
\end{eqnarray}
respectively. Note that the leading spin contribution to ${\cal
  A}_\parallel$ vanishes because of G-parity (for massless quarks).

The DAs ${\cal V}_\perp$ and ${\cal A}_\perp$, on the other hand,
correspond to the projection  $\gamma_\perp G_{\cdot\perp}$ and thus
do not describe states with a definite projection of the quark spins
onto the light-ray $z_\mu$. We separate the different quark spin
projections with the aid of the auxiliary amplitudes ${\cal
  H}^{\uparrow\downarrow}$ and ${\cal H}^{\downarrow\uparrow}$ defined as
\begin{eqnarray}
\langle 0 | \bar u(z) ig\widetilde{G}_{\alpha\beta}(vz) \gamma_\cdot
\gamma_\mu\gamma_* d(-z) | \pi^-\rangle & = & f_\pi m_\pi^2
\left( p_\beta g^\perp_{\alpha\mu} - p_\alpha
  g^\perp_{\beta\mu} \right)
{\cal H}^{\uparrow\downarrow}(v,pz),\nonumber\\
\langle 0 | \bar u(z) ig\widetilde{G}_{\alpha\beta}(vz) \gamma_*
\gamma_\mu\gamma_\cdot d(-z) | \pi^-\rangle & = & f_\pi m_\pi^2
\left( p_\beta g^\perp_{\alpha\mu} - p_\alpha
  g^\perp_{\beta\mu} \right)
{\cal H}^{\downarrow\uparrow}(v,pz).
\end{eqnarray}
The original distributions ${\cal A}_\perp$ and ${\cal V}_\perp$ are
then given by
\begin{eqnarray}
{\cal V}_\perp(\underline{\alpha}) & = & -\frac{1}{2}\left( {\cal H}^{
\uparrow\downarrow}(\underline{\alpha}) + {\cal
H}^{\downarrow\uparrow}(\underline{\alpha})\right),\nonumber\\
{\cal A}_\perp(\underline{\alpha}) & = & \phantom{-}
\frac{1}{2}\left( {\cal H}^{
\uparrow\downarrow}(\underline{\alpha}) - {\cal
H}^{\downarrow\uparrow}(\underline{\alpha})\right).\label{eq:original}
\end{eqnarray}
${\cal H}^{\uparrow\downarrow}$ and ${\cal H}^{\downarrow\uparrow}$
have a simple expansion in terms of Appell polynomials, to wit:
\begin{eqnarray}
{\cal H}^{\uparrow\downarrow}(\underline{\alpha}) & = &
60\alpha_u\alpha_g^2 \left[ h_{00} + h_{01}(\alpha_g-3\alpha_d) +
  h_{10}\left(\alpha_g-\frac{3}{2}\alpha_u\right)\right],\nonumber\\
{\cal H}^{\downarrow\uparrow}(\underline{\alpha}) & = &
60\alpha_d\alpha_g^2 \left[ h_{00} +h_{01}(\alpha_g-3\alpha_u) +
  h_{10}\left(\alpha_g-\frac{3}{2}\alpha_d\right)\right],
\end{eqnarray}
where we have taken into account the symmetry properties, i.e.\ $${\cal
  H}^{\uparrow\downarrow}(\alpha_d,\alpha_u) \equiv {\cal
  H}^{\downarrow\uparrow}(\alpha_u,\alpha_d).$$
{}From (\ref{eq:original}), the following relations can be derived 
 immediately:
\begin{eqnarray}
{\cal V}_\perp(\underline{\alpha}) & = &
 -30 \alpha_g^2\left[ h_{00}(1-\alpha_g)
                    +h_{01}\Big[\alpha_g(1-\alpha_g)-6\alpha_u\alpha_d\Big]
\right.\nonumber\\
& & \left.
                    +h_{10}\Big[\alpha_g(1-\alpha_g)-\frac{3}{2}(\alpha_u^2
                               +\alpha_d^2)\Big]\right],
\nonumber\\
{\cal A}_\perp(\underline{\alpha}) & = &
 30 \alpha_g^2(\alpha_u-\alpha_d)\left[ h_{00}
                    +h_{01}\alpha_g
                    +\frac{1}{2}\,h_{10}(5\alpha_g-3)
                                 \right].
\end{eqnarray}
The DAs ${\cal V}_{\perp,\parallel}$ and ${\cal A}_{\perp,\parallel}$
depend, to next-to-leading accuracy in the conformal spin, on a total
of six parameters: $v_{00}$ and $h_{00}$ of leading conformal spin and
$v_{10}$, $a_{10}$, $h_{10}$ and $h_{01}$ of NLO conformal spin. Our next
task is to relate these
parameters to independent local matrix elements. Defining
\begin{equation}
\langle 0 | \bar u\gamma_\xi
ig\widetilde{G}_{\xi\alpha}d|\pi^-\rangle = f_\pi m_\pi^2 \eta_4
P_\alpha,
\end{equation}
which is equivalent to
$$
\langle 0 | \bar u \gamma_\alpha ig\widetilde{G}_{\mu\nu}
d|\pi^-\rangle = -\frac{1}{3}\, f_\pi m_\pi^2 \eta_4 (P_\mu
g_{\nu\alpha} - P_\nu g_{\mu\alpha}),
$$
it follows
\begin{equation}
h_{00} = v_{00} = -\frac{1}{3}\,\eta_4.
\end{equation}
$\eta_4$ is scale-dependent with
$$
\eta_4(Q^2) = L^{\gamma_4^\eta/b}\, \eta_4(\mu^2), \quad \gamma_4^\eta =
\frac{8}{3}\, C_F.
$$

To NLO in conformal spin, beyond the matrix elements already defined
above, we need only one more matrix element of a conformal quark-gluon
operator:
\begin{eqnarray}
\lefteqn{\langle 0 | \bar u [iD_\mu,ig\widetilde{G}_{\nu\xi}] \gamma_\xi d -
\frac{4}{9}\, i\partial_\mu \bar u i g
\widetilde{G}_{\nu\xi}\gamma_\xi d | \pi^-\rangle\
=}\hspace*{0.5cm}\nonumber\\
& = & f_\pi m_\pi^2 \eta_4 \omega_4 \left(P_\mu P_\nu - \frac{1}{4}\, m_\pi^2
g_{\mu\nu}\right) + O({\rm twist\ 5}).
\end{eqnarray}
The scale-dependence of $\omega_4$ is given by
$$
\omega_4(Q^2) = L^{\gamma_4^\omega/b}\, \omega_4(\mu^2), \quad 
\gamma_4^\omega = \frac{10}{3}\, C_A - \frac{8}{3}\, C_F.
$$
Numerical values for $\eta_4$ and $\omega_4$ were calculated from QCD sum rules
\cite{novikov,BF90} and are collected in Table~\ref{tab:3}. In the
notation of Ref.~\cite{BF90}, $\delta^2 \equiv m_\pi^2 \eta_4$, 
$\epsilon\equiv 21/8\, \omega_4$.

The procedure how to relate $v_{10}$, $a_{10}$, $h_{10}$ and $h_{01}$ 
to local matrix elements is
described in detail in Ref.~\cite{HT98}, so that we mention only the
essentials. Three of the four necessary relations follow from an
analysis of various light-cone projections of the matrix elements of
the operators
\begin{eqnarray*}
 O^{(1)}_{\alpha\beta\mu\nu} & = & -\bar u (i\derleft_\beta
 g\wt{G}_{\mu\nu} + g\wt{G}_{\mu\nu}
 i\derright_\beta)\gamma_\alpha d,\\
 O^{(2)}_{\alpha\beta\mu\nu} & = & \bar u (-i\derleft_\beta
 g{G}_{\mu\nu} + g{G}_{\mu\nu}
 i\derright_\beta)\gamma_\alpha\gamma_5 d.
\end{eqnarray*}
The fourth relation can be derived from the operator identity
\begin{eqnarray*}
\frac{4}{5} \partial_\mu E_{\mu\alpha\beta} & = & -12 i \bar u
\gamma_\rho\gamma_5
\left\{G_{\rho\beta} \derright_\alpha - \derleft_\alpha G_{\rho\beta} +
  (\alpha\leftrightarrow \beta) \right\} d
-4\partial_\rho \bar u (\gamma_\beta\wt{G}_{\alpha\rho} +
\gamma_{\alpha} \wt{G}_{\beta\rho} ) d\\
& & {} - \frac{8}{3}\,
\partial_\beta \bar u \gamma_\sigma \wt{G}_{\sigma\alpha} d
- \frac{8}{3}\,\partial_\alpha \bar u \gamma_\sigma
\wt{G}_{\sigma\beta} d
+\frac{28}{3}\, g_{\alpha\beta} \partial_\rho \bar u \gamma_\sigma\gamma_5
\wt{G}_{\sigma\rho} d,
\end{eqnarray*}
where
$$E_{\mu\alpha\beta} =\left[
\frac{15}{2}\bar u \gamma_\mu\gamma_5 \deriv_{\alpha} \deriv_{\beta}  d
-\frac{3}{2} \partial_{\alpha} \partial_{\beta}\bar u \gamma_\mu\gamma_5  d
  - {\rm traces}\right]_{\rm symmetrized}
$$
is a leading twist 2 conformal operator. The matrix element of this
operator is proportional to the Gegenbauer moment $a_2$ of the twist 2
DA and brings in a dynamical mass
correction in the twist 4 DAs. This is precisely the effect mentioned
in the introduction: total derivatives of twist 2
operators enter higher twist DAs. 

The final results for the NLO parameters read:
\begin{eqnarray}
a_{10} & = & \frac{21}{8}\,
\eta_4\omega_4 -\frac{9}{20}\, a_2 ,\nonumber\\
v_{10} & = & \frac{21}{8}\, \eta_4 \omega_4,\nonumber\\
h_{01} & = & \frac{7}{4}\, \eta_4
\omega_4 -\frac{3}{20}\, a_2 ,\nonumber\\
h_{10} & = & \frac{7}{2}\,\eta_4\omega_4 + \frac{3}{20}\, a_2.
\end{eqnarray}
For $a_2\to 0$, these results agree with those obtained in \cite{BF90}.

\FIGURE[b]{\epsfig{file=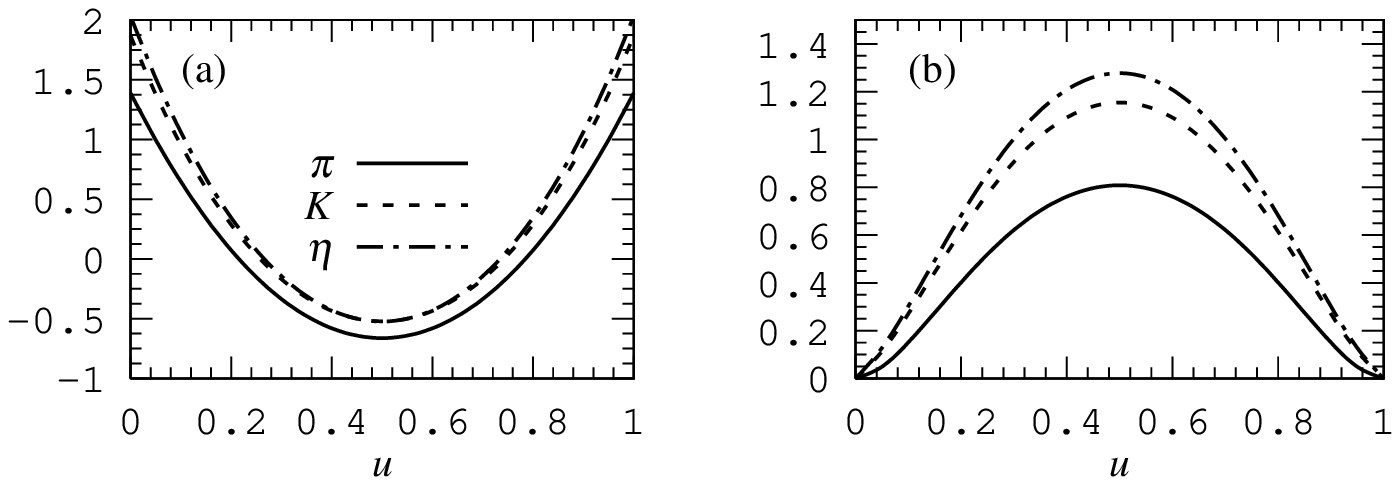}\caption{The two particle DAs of twist
    4: $g_P(u)$ (a) and $\mathbb A$ (b) for $P=\pi,\,K,\,\eta$.}
\label{fig:2}}

We are now in the position to derive expressions also for the
remaining two DAs of twist 4, $g_\pi$ and $\mathbb A$. From the
operator relations (\ref{eq:oprel1}) and (\ref{eq:oprel2}), one
obtains
\begin{eqnarray}
g_\pi(u) & = & 2\phi_p(u) - \phi_\pi(u) + \frac{d\phantom{u}}{du}
\int_0^u d\alpha_d \int_0^{\bar u} d\alpha_u \,\frac{2}{\alpha_g}
\left( {\cal A}_\parallel(\underline{\alpha}) - 2 {\cal
    A}_\perp(\underline{\alpha}) \right),\label{eq:gpi}\\
{\mathbb A}(u) & = & 12\int_0^u dv \int_0^v dw \,
(g_\pi(w)-\phi_\pi(w)) - 2 \int_0^u dv (2v-1) (\phi_\pi(v) + g_\pi(v))
\\
& & + \int_0^u d\alpha_d \int_0^{\bar u} d\alpha_u \,\frac{4}{\alpha_g^2}
\, (\alpha_d-\alpha_u - \xi) ( 2 {\cal A}_\perp(\underline{\alpha}) -
{\cal A}_\parallel(\underline{\alpha}) ).
\end{eqnarray}
$g_\pi$ corresponds to a definite quark spin projection and thus has a
simple expansion in Gegenbauer polynomials $C_{n}^{1/2}$:
\begin{equation}
g_\pi(u) = \sum_{i=0}^\infty g_{2i} C_{2i}^{1/2}(\xi).
\end{equation}
{}From (\ref{eq:gpi}), one finds:
\begin{eqnarray*}
g_0 & = & 1,\\
g_2 & = & 1 + \frac{18}{7}\, a_2 + 60\eta_3 + \frac{20}{3}\, \eta_4,\\
g_4 & = & -\frac{9}{28}\, a_2 - 6\eta_3\omega_3,
\end{eqnarray*}
where we neglect terms of $O(m_u+m_d)$ induced by $\phi_p$, as $g_\pi$
itself enters the matrix element of the axialvector current already as
$O(m_\pi^2) = O(m_u+m_d)$, cf.\ (\ref{eq:2.8}). 

The expansion of $\mathbb A$ is not that straightforward and
involves logarithms:
\begin{eqnarray*}
\lefteqn{{\mathbb A}(u)\ =\ 6u\bar u \left\{ \frac{16}{15} + \frac{24}{35} \,
  a_2 + 20 \eta_3 + \frac{20}{9} \,\eta_4 \right.}\\
& &\left. + \left( -\frac{1}{15} + \frac{1}{16}\, - \frac{7}{27}\, \eta_3
  \omega_3 - \frac{10}{27}\, \eta_4 \right) C_2^{3/2}(\xi) + \left
  ( -\frac{11}{210} \, a_2 - \frac{4}{135}\, \eta_3\omega_3 \right)
  C_4^{3/2}(\xi) \right\} \\
& & {}+ \left(-\frac{18}{5}\, a_2 + 21\eta_4\omega_4 \right) \left\{ 2
  u^3 (10-15 u + 6 u^2)\ln u + 2\bar u^3 (10-15\bar u + 6 \bar u^2)
  \ln\bar u\right.\\
& & \hspace*{4.3cm}\left. + u \bar u (2 + 13u\bar u)\right\}.
\end{eqnarray*}
\TABLE{
\begin{tabular}{|c|ccc|}
\hline
& $\pi$ & $K$ & $\eta$\\ \hline
$a_2$ & 0.44 & 0.2 & 0.2\\
$\eta_3$ & 0.015 & 0.015 & 0.013\\
$\omega_3$ &  $-3$ & $-3$ & $-3$\\
$\eta_4$ & 10 & 0.6 & 0.5\\
$\omega_4$ & 0.2 & 0.2 & 0.2\\ \hline
\end{tabular}
\caption{Input parameters for twist 4  DAs, calculated from QCD sum rules.
The accuracy is about 30\%. Renormalization scale is
1~GeV.}
\label{tab:3}}
In Fig.~\ref{fig:2} we plot $m_P^2 g_P$ and $m_P^2 {\mathbb A}$ for
the mesons $P=\pi,K,\eta$. 
Whereas the DA $m_P^2 g_P$ is not too different for the three different
mesons, the impact of meson-mass corrections on $m_P^2 {\mathbb A}$ is
more noteworthy. For the area under the curve, i.e.\ the overall
normalization $N_P = \int_0^1 du\, m_P^2 {\mathbb A}(u)$, we find:
$$
N_\pi = 0.47, \quad N_K = 0.70, \quad N_\eta = 0.77.
$$
The result for $N_\eta$ is essentially independent of the precise
value of $f_\eta^8$.
The impact of meson-mass corrections is thus rather profond for the DA
$\mathbb A$.
Likewise the change in normalization of
$\phi_\sigma$, this shift in $\mathbb A$ may have a noticeable effect
on the $B\to K$ form factors calculated from QCD sum rules on the light-cone.

\section{Summary and Conclusions}
\setcounter{equation}{0}

In this paper we have studied the twist 3 and 4 two and three particle
DAs of pseudoscalar nonsinglet mesons in QCD and expressed them in a
model-independent way by a minimal number of nonperturbative
parameters. The work presented here is an extension of the paper 
\cite{BF90} on
$\pi$ DAs and is in particular concerned with corrections in the meson
mass. The one ingredient in our approach is the use of the QCD
equations of motion, which allow one to express dynamically dependent
DAs in terms of independent ones. The other ingredient is conformal
expansion which makes it possible to separate transverse and
longitudinal variables in the wave functions, the former ones being
governed by renormalization group equations, the latter ones being
described in terms of irreducible representations of the corresponding
symmetry group.

The analysis of twist 4 DAs is complicated by the
fact that the twist 4 terms are of different origin: there are, first,
``intrinsic'' twist 4 corrections from matrix elements of twist 4
operators. There are, second, admixtures of matrix elements of twist 3
operators, as the counting of twist in terms of ``good'' and ``bad''
projections on light-cone coordinates does not exactly match the
definition of twist as ``dimension minus spin'' of an operator. There
are, third, meson-mass corrections, which one may term  kinematical
corrections, that come, on the one hand,  from the subtraction of traces in the
leading twist operators and, on the other hand,
from higher twist operators containing
total derivatives of twist 2 operators.
Meson-mass corrections of the first kind are formally
analogous to Nachtmann corrections in inclusive processes, while
the contribution of  operators with total derivatives
is a specific new feature in exclusive
processes, which makes the structure of these corrections much more
complex.

Our final results are collected in Secs.\ 5 and 6. In contrast to the
case of vector mesons whose mass is finite also in the chiral limit,
pseudoscalar octet mesons bring in the  complication that
their mass squared depends linearly on the quark masses, so that for
consistency one also has to take into account terms of
$O((m_u+m_d)^2/m_\pi^2) \sim O(m_\pi^2)$. The effect of meson-mass
corrections is noticeable for twist 3 DAs and in particular for the twist 4 DA
$\mathbb A$, whose normalization increases by $\sim 50$\%, when
comparing the $\pi$ with the $K$. 

We hope that our results will contribute to a better understanding of
SU(3) breaking effects in hard exclusive processes and in particular to
the investigation of $B$ and $B_s$ decay form factors into
$\pi$, $K$ and $\eta$ mesons from QCD sum rules on the light-cone.

\acknowledgments

The author is supported by DFG through a Heisenberg fellowship.

\end{document}